# 5

# Leveraging SDN for The 5G Networks: Trends, Prospects and Challenges

Akram Hakiri and Pascal Berthou

*CNRS, LAAS, 7 Avenue du colonel Roche, F-31031 Toulouse, France*

*Univ de Toulouse, UPS, LAAS, F-31031 Toulouse, France*

*firstname.lastname@laas.fr*

## Abstract

Today 4G mobile systems are evolving to provide IP connectivity for diverse applications and services up to 1Gbps. They are designed to optimize the network performance, improve cost efficiency and facilitate the uptake of mass market IP-based services. Nevertheless, the growing demand and the diverse patterns of mobile traffic place an increasing strain on cellular networks. To cater to the large volumes of traffic delivered by the new services and applications, the future 5G network will provide the fundamental infrastructure for billions of new devices with less predictable traffic patterns will join the network. The 5G technology is presently in its early research stages, so researches are currently underway exploring different architectural paths to address their key drivers. SDN techniques have been seen as promising enablers for this vision of carrier networks, which will likely play a crucial role in the design of 5G wireless networks. A critical understanding of this emerging paradigm is necessary to address the multiple challenges of the future SDN-enabled 5G technology. To address this requirement, a survey the emerging trends and prospects, followed by in-depth discussion of major challenges in this area are discussed.

## 5.1 Introduction

Mobile and wireless connectivity have made tremendous growth during the last decade. Today, the 3G/4G mobile wireless systems are becoming in the ground to provide connectivity through IP core network (i.e., Evolved Packet Core (EPC)). They also focus towards providing seamless connection to cellular networks such as 3G, LTE, WLAN and Bluetooth. The 5G (fifth Generation) is being seen as user-centric concept instead of operator-centric as in 3G or service-centric as seen for 4G. Mobile terminals will be able to combine multiple flows incoming from different technologies. Multimode mobile terminals have been seen towards the 4G cellular network. They aim to provide single user terminal that can cooperate in different wireless networks and overcome the design problem of power-consumption and cost old mobile terminals.

The Open Wireless Architecture (OWR) [1] is targeted to support multiple existing wireless air interfaces as well as future wireless communication standard in an open architecture platform. Nevertheless, the growing demand and the diverse patterns of mobile traffic place an increasing strain on cellular networks. To cater to the large volumes of traffic delivered by the new services and applications, the future fifth generation 5G of wireless/mobile broadband [2] network will provide the fundamental infrastructure for billions of new devices with less predictable traffic patterns will join the network.

The 5G wireless network should enable the development and exploitation of massive capacity and massive connectivity of complex and powerful heterogeneous infrastructures. Accordingly, the network should be capable of handling the complex context of operations to support the increasingly diverse set of new and yet unforeseen services, users and applications (i.e., including smart cities, mobile industrial automation, vehicle connectivity, machine-to-machine (M2M) modules, video surveillance, etc.), all with extremely diverging requirements, which will push mobile network performance and capabilities to their extremes. Additionally, it should provide flexible and scalable use of all available non-contiguous spectrums (e.g., further LTE enhancements to support small cells (Non-Orthogonal Multiple Access (NOMA) [3], Future Radio Access (FRA)) for wildly different network deployment scenarios, in an energy efficient and secure manner).

To address these key challenges, there is a need to enhance the future networks through intelligence, to proceed to successful deployment and realization of a powerful wireless world. Principals of virtual network management and operation, network function virtualization (NFV), and Software-Defined Networking (SDN) [4] are redefining the network architecture to support the new requirements of a new eco-system in the future. SDN techniques have been seen as promising enablers for this vision of carrier cloud, which will likely play a crucial role in the design of 5G wireless networks. Accordingly, the future SND-enabled 5G communications have to properly address key challenges and requirements driven by multiple society, users, and operators, which would give them greater freedom to balance operational parameters, such as network resilience, service performance and Quality of Experience (QoE).

## 5.2 Evolution of the Wireless Communication towards the 5G

Figure 5. 1 depicts an overview of the wireless world towards the 5G. The Figure gives a multi-dimensional overview of significant design challenges that 5G technology will face to simultaneously meet the future services, as to achieve cost-effective resource provisioning and ecosystem, built with novel technologies such as SDN and network virtualization.

5.2.1 Evolution of the wireless world

The mobile communication system has evolved through the first generation (1G), to the second generation (2G) and third generation (3G) through the 4G or Long-Term Evolution-Advanced (LTE-A) of mobile/cellular communications, with the typical service improvement and cost efficiency for each generation. For example, 1G (i.e., Advanced Mobile Phone System (AMPS)) and 2G (i.e., GSM and GPRS) were designed for circuit switched voice application. 3G (i.e., UMTS) and 4G (i.e., LTE-Advanced) were developed for packet switched services including multimedia, wide-band data and mobile Internet services. Meanwhile, there has been the introduction of other local, metropolitan, and wide area wireless/cellular technologies such as microcells, Femtocell, Pico-cell, small cells, etc.

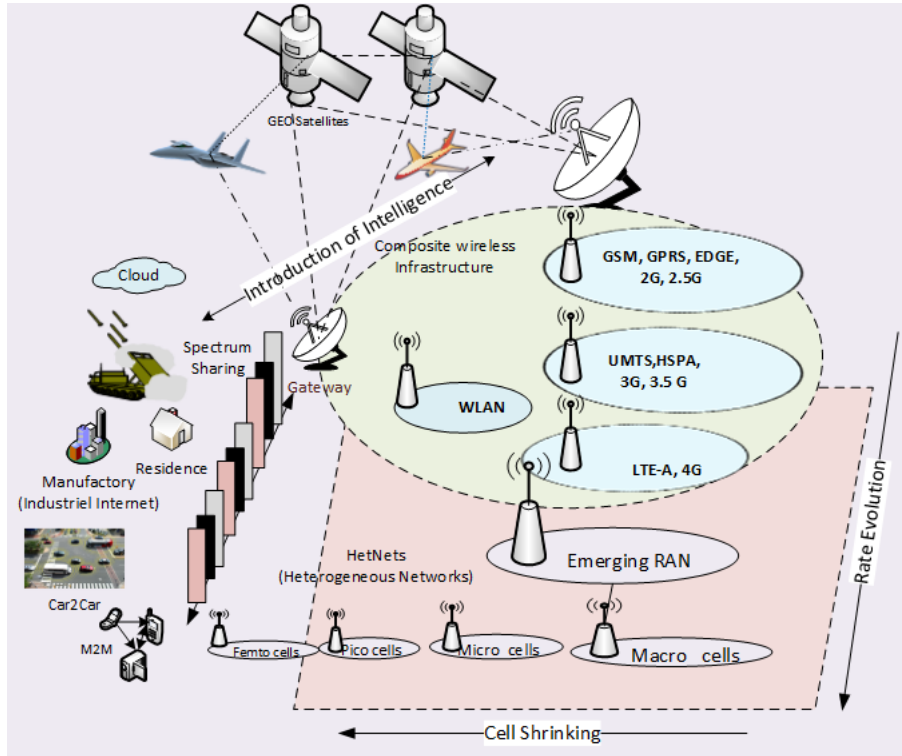

Figure 5. 1: A View of the Wireless World

The other evolution, which has emerged in the past decade, aimed to exploit heterogeneous wireless communication, comprised the wireless access infrastructure in both licensed and unlicensed part of the wireless spectrum. It was intended to interconnect cellular system to wireless access networks (i.e., WLAN, WiMAX, etc.) to improve the service delivery and application provisioning end-to-end. Thus, the network was composed of different mixed types of infrastructures, forming heterogeneous networks (HetNets) [5]. Driven by both technical and economic incentives, the proliferation of HetNets had offered opportunities to satisfy users and applications in terms of their capabilities to support the new services.

Another important direction, which is expected to characterize beyond the 4G and 5G wireless networks, concerns the deployment of the application driven networks. Application-driven networks consist of interconnecting end-user devices, M2M (Machine to Machine) modules, and several machines, sensors, and actuators, so called IoT (Internet of Things) with billions of objects connected to the internet for supplying big data applications. In parallel, in recent years the introduction and deployment of cloud-based concepts have emerged as an important solution offering enterprises a potentially cost-effective business model. For example, mobile users can use cloud-connected devices through public and private Mobile Personal Grid (MPG). Given the dynamic needs and supply of the network resource with rich resources available in the cloud, mobile users can benefit from resource virtualization to accommodate the different requirements as mobile devices move around within the mobile cloud. Furthermore, the integration of satellites in the future 5G networks reveals many challenges to support flexible, programmable and secure infrastructure. The intersection of the cloud, satellites, Big Data, M2M, and 5G will bring about an exciting new automated future.

5G networks will not be based on routing and switching technologies anymore. They will be open, more flexible, able to support HetNets, and able to evolve more easily than the traditional networks. They will be able to provide convergent network communication across multi-technologies networks (e.g., packet and optical networks), and provide open communication system to cooperate with satellite systems, cellular networks, clouds and data-centers, home gateways, and many more open networks and devices. Additionally, 5G systems will be autonomous and sufficiently able to adapt their behavior depending in the user's requirements to handle application-driven networks in dynamic and versatile environments. Security, resiliency, robustness and data integrity will be a key requirement of future networks.

## 5.3 Software-Defined Networks

Introduction of intelligence towards 5G can address the complexity of Heterogeneous Networks (HetNets) by specifying and providing flexible solutions to cater for network heterogeneity. Software-Defined Networking (SDN) has emerged as a new intelligent architecture for network programmability. The primary idea behind SDN is to move the control plane outside the switches and enable external control of data through a logical software entity called controller. SDN provides simple abstractions to describe the components, the functions they provide, and the protocol to manage the forwarding plane from a remote controller via a secure channel. This abstraction captures the common requirements of forwarding tables for a majority of switches and their flow tables. This centralized up-to-date view makes the controller suitable to perform network management functions while allowing easy modification of the network behavior through the centralized control plane.

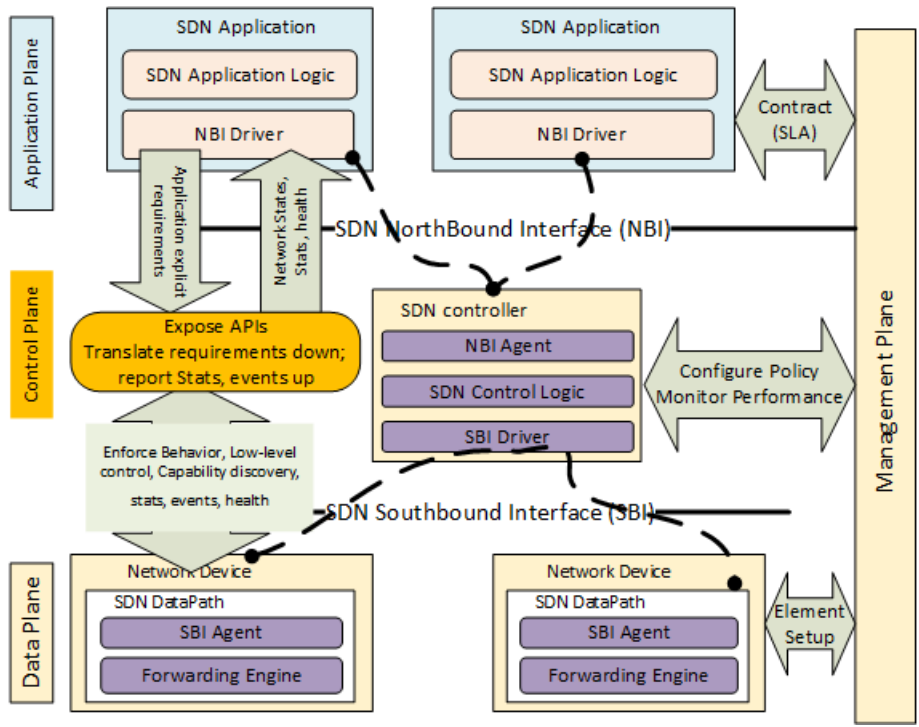

Figure 5. 2: Reference Architecture of Software-Defined Networking

**Figure 5. 2** depicts the overall SDN architecture. The SDN community has adopted a number of northbound interfaces (i.e., between the control plane and applications) that provide higher level abstractions to program various network-level services and applications at the control plane. For the southbound interface (i.e., between the control plane and network devices), the OpenFlow standard [6] has emerged as the dominant technology. For instance, consider the operation of an Ethernet switch. From the functional point of view, Ethernet switches can be divided into a data plane and a control plane. The data plane represents a forwarding table according to which incoming packets to an Ethernet switch are forwarded. Forwarding tables consist of entries which tell to which output port the received Ethernet frames should be sent. Populating of forwarding table with these entries is the task of the control plane. The control plane is a set of actions exerted on the received Ethernet frames to decide their destination ports. In order to quickly perform frame processing, these actions are implemented in hardware together with the forwarding table.

SDN makes it possible to manage the entire network through intelligent orchestration and provisioning systems. Thus it allows on-demand resource allocation, self-service provisioning, truly virtualized networking, and secures cloud services. Thus, the static network can evolve into an extensible vendor-independent service delivery platform capable of responding rapidly to changing business, end-user, and market needs, which greatly simplifies the network design and operation. Consequently, the devices themselves no longer need to understand and process thousands of protocol standards but merely accept instructions from the SDN controllers.

The value of SDN in 5G wireless networks lies specifically in its ability to provide new capabilities like network virtualization, automating and creating new services on top of the virtualized resources, in secure and trusted networks. Also, SDN enables the separation of the control logic from vendor-specific hardware to open and vendor-neutral software controllers. Thus, it enables implementing routing and data processing functions of wireless infrastructure into software packages in general purpose computer or even in the cloud.

## 5.4 Network Function Virtualization

One of the most interesting complementary technology of SDN which has the potential to dramatically impact the future 5G networking and how to refactor the architecture of legacy networks, is virtualizing as many network functions as possible, so called Network Function Virtualization (NFV). The aim of NFV is to virtualize (known also as network softwarization) a set of network functions by deploying them into software packages, which can assembled and chained to create the same services provided legacy networks. It is possible for example to deploy a virtualized Session Border Controller (SBC) [7] in order to protect the network infrastructure more easily than installing the conventional complex and expensive network equipment's. The concept of NFV is inherited from the classical server virtualization that could by installing multiple virtual machines running different operating systems, software and processes.

Traditionally, network operators had always preferred the use dedicated high available black-box network equipment's to deploy their networks. However, this old approach inevitably leads to long time-to-market (CapEx) and requires a competitive staff (OpEx) to deploy and run them. As depicted in **Figure 5. 3Erreur ! Source du renvoi introuvable.**, the NFV technology aims to build an end-to-end infrastructure and enable the consolidation of many heterogeneous network devices by moving network functions from dedicated hardware onto general purpose computing/storage platforms such as servers. The network functions are implemented in software packages that can be deployed in virtualized infrastructure, which will allow for new flexibilities in operating and managing mobile networks.

Another important topic in 5G carrier-grade mobile networks which may be improved by implementing NFV in cloud infrastructures is resilience. Implementing network functions in data centers allows transparent migration between either virtual machines or real machines. Furthermore, implementing mobile network functions in data centers will enable more flexibility in terms of resource management, assignment, and scaling. This impact the development of eco-systems and energy efficiency of networks, as over-provisioning can be avoid by only using the necessary amount of resources.

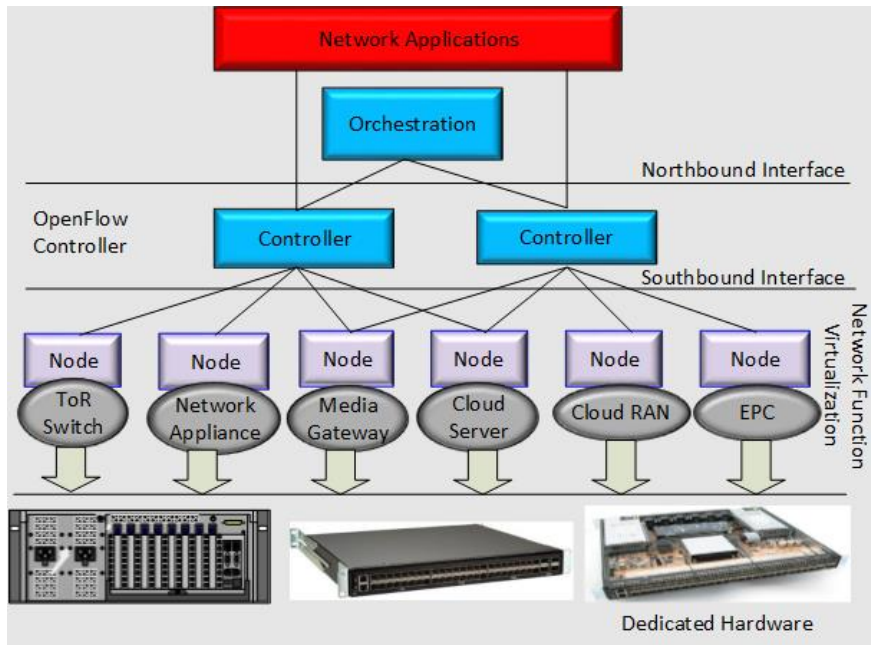

Figure 5. 3: Network Function Virtualization

NFV is currently discussed in the context of virtualizing the core network as well as centralizing the base band processing within Radio Access Networks (RAN). Examples of mobile network virtualization are used for Cloud-RAN (C-RAN). C-RAN can use virtualized software modules running in different virtual machines. Additionally, enhancing NFV with SDN may offload the centralized location within networks nodes which require high-performance connections between Radio Access Point (RAP) and data centers. Decentralizing these connections with SDN will enable managing heterogeneous network nodes (i.e., Pico-cell, macro-cell, etc.) and heterogeneous back-haul connectivity such as fiber, wireless, etc.

Another concept that received a lot of attention with the evolution of SDN and NFV is the Network Service Chaining (NSC) [8]. The NSC aimed to help carrier-grade to provide continuous delivery of services based on dynamic network function orchestration and automated deployment mechanisms to improve operational efficiency. Because SDN moves the management functions out of the hardware and places them in controller software running in general purpose server, and NFV moves network functions out of hardware and puts them onto software, too, building service chaining no longer requires hardware, so there is no need for over-provisioning since additional servers can be added when needed.

One example of an increasingly complex network platform is the 3GPP Evolved Packet Core (EPC), which requires multiple functions (e.g., Network Address Translation (NAT), service access policing for VPN, video platforms and VoIP, infrastructure firewall protection, etc.) typically installed in independent boxes. Carrier-grade networks should define statically- One example of an increasingly complex network platform is the 3GPP Evolved Packet Core (EPC), which requires multiple functions (e.g., Network Address Translation (NAT), service access policing for VPN, video platforms and VoIP, infrastructure firewall protection, etc.) typically installed in independent boxes. Carrier-grade networks should define statically-provisioned service chains for customer traffic crosses several middle boxes. In the future 5G networks carrier-grade will not use monolithic, closed, and mainframe-like boxes to provide a single service. SDN and NFV-driven service chaining can improve flexible allocation, orchestration and management of cross-layer (L2-L7) network functions and services and provides the substrate for dynamic network service chains.

## 5.5 Information-Centric Networking

The Information-Centric Networking (ICN) is a novel network architecture that is receiving a lot of attention in the 5G networks. ICN consists of new communication model that revolve around the production, consumption and matching users with content, in-network caching and content-based service differentiation, instead of communication channels between hosts. ICN pushes many design principals from the Web to the network architecture by centering on what is relevant to the user and not where the content is located in the network. So, ICN manages contents and names to ensure their uniqueness in the network (i.e., because data are routed based on their names). The ICN communication model allows built-in native features aiming at optimizing and simplifying future content delivery architecture. The service providers should prepare their infrastructure capabilities to support efficient multicast data delivery as well as provide seamless mobile connectivity so users can move and the network can continue delivering data packets without interruption.

Typically, the ICN deployment schemes can be classified into three categories: (1) ICN over IP, which encapsulate ICN protocol data in IP (or UDP/TCP) packets or take ICN protocol information using IP options; (2) ICN over L2, which completely replaces the IP layer and directly uses the data link protocol (such as PPP, Ethernet, IEEE 802.x) to deliver data between neighbors; and (3) ICN over virtualized network, which exploits network virtualization technologies, such as SDN, to implement ICNs. Although these schemes have advantages and disadvantages, most works on ICN implementations focus on how to implement a particular ICN architecture. However, different ICN architectures employ different transmission techniques and packet format, which is not easy for the co-existence and inter-operability of different ICNs. SDN and NFV are amazing approaches for improving the integration of ICN in the 5G network without deploying new ICN capable hardware.

Levering SDN for ICN requires a unified framework to facilitate the implementation and interoperability among different ICN architectures [9]. Such a content-centric framework should provide users network access to remote named-resources, rather than to remote hosts [10]. The integration of ICN in the 5G network includes storage and execution capabilities to evolve the network from a dumb pipe transport towards added value intelligent network. Introduction of intelligence in ICN architecture improves the flexibility and scalability of content-naming as well as enhances the performance of the QoS in the network. Intelligent ICN can also made it feasible to integrate mobile radio aware ICN on the 5G networks.

Although supporting ICN-enabled SDN allows transforming the current network model into simplified, programmable and generic one, ICN still faces a number of challenges to its realization including routing computation, path labeling to discover the network topology and locate data in the network, and routes assignments to route requests for data objects. Moreover, since ICN information should be inserted in each packet, the fragmentation of packets limits processing cost of the network resources. Mechanisms for caching objects in the network along the path need more investigation to deliver them more rapidly to an increasing number of users. Distribution of storage capabilities across the path with more elaborated content-routing algorithms are open issues, so researchers have to cope with a proliferation of new and complex application-contents and services, many of which are unknown today.

## 5.6 Mobile and Wireless Networks

The design of the future 5G systems should efficiently support a multitude of diverse services and introduces new methods making the network application-service aware. The future 5G network architecture should be highly flexible for supporting traditional use cases as well as easy integration of future one's. Additionally, the 5G network will be able of handling user-mobility, while the terminals will make the final choice among different access networks transparently. Mobile terminals will also hold intelligent components to make choice of the best technology to connect, with respect of the constraints and dynamically change the current access technology while guaranteeing the end-to-end-connectivity.

### 5.6.1 Mobility Management

A key trend relates to mobility, as broadband mobile is expected to growth in the next decade. The future will encompass 1000 times more connected mobile device in the horizon 2020, all with different QoS requirements, which will interconnect to all kinds of heterogeneous and customized Internet based services and application. Accordingly, these developments demand rethinking about the network design, which leads to ask about the advantages of SDN in the most common wireless networking scenarios. It is also important to understand what the key challenges that exist in this realm are and how they can be addressed.

Presently, there is not much discussion regarding the mobility support in SDN. Software Defined Wireless Networking (SDWN) would be a SDN technology for wireless/broadband networks that provides radio resource management, mobility management and routing [11]. SDWN infrastructure can support composition to combine results of multiple modules into a single set of packet-handling rules. For example, novel mobility management protocols should be provided to maintain session continuity from the application's perspective and network connectivity through dynamic channel configuration. Furthermore, mobility modules should be able to provide rapid client re-association, load balancing, and policy management (i.e., charging, QoS, authentication, authorization, etc.)

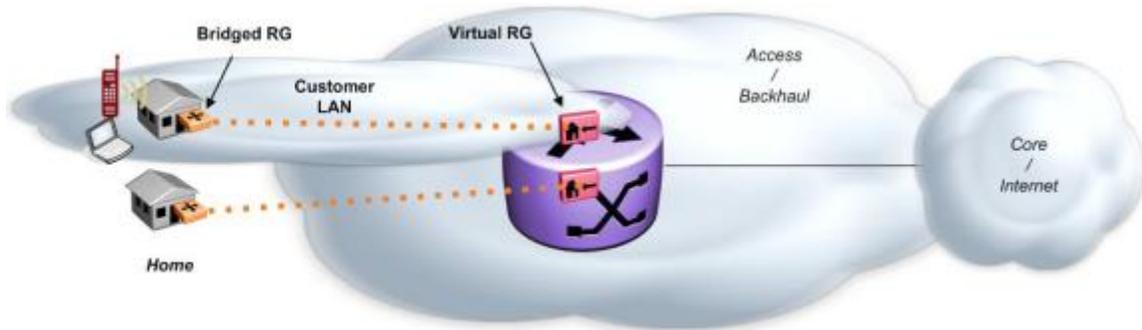

Figure 5. 4: Orange Cloud Box Virtualized Residential Gateway

Another important key challenge in wireless/broadband networks concerns multi-homing. Multi-homing means the attachment of end-host to multiple networks at the same time, so users could freely move between wireless infrastructures while also supporting the provider. This approach would emerge by applying SDN capabilities to relay between the home network and edge networks. The future wireless/broadband system can be envisioned as a world in which mobile devices can moves seamlessly between the wireless infrastructures, in trust and secure manner. For example, as shown in **Figure 5. 4** in home networks a virtualized residential gateway can improve service delivery between the core home network and the network-enabled devices. The target architecture emerges by applying SDN and NFV between the home gateway and the access network, moving most of the gateway functionality to a virtualized execution environment.

### 5.6.2 Ubiquitous Connectivity

A part of the future wireless networks, end-users will need to communication with each other and with a surrounding objects and machines, e.g., sensors embedded in objects. **Figure 5. 5**Figure 5. **5** shows how the cellular network will be completed by interaction with network topologies, including M2M, which completed with user/device-to-user/device communication at different level of cooperation and coordination between different nodes.

Taking into account all kinds of interactions of these ubiquitous systems, will increasingly expend the network infrastructure, which will include new data services and applications, e.g., smart-phones and tablet with powerful multimedia capabilities or even connected things surrounding the environment, such as building, roads or even in car-to-car communication. Accordingly, essential design criteria to fulfill the requirements of future 5G systems are fairness between users over the covered area of ubiquitous systems, latency reduce, increase reliability, energy efficiency, and enhance QoS and QoE requirements originated from heterogeneous applications and services.

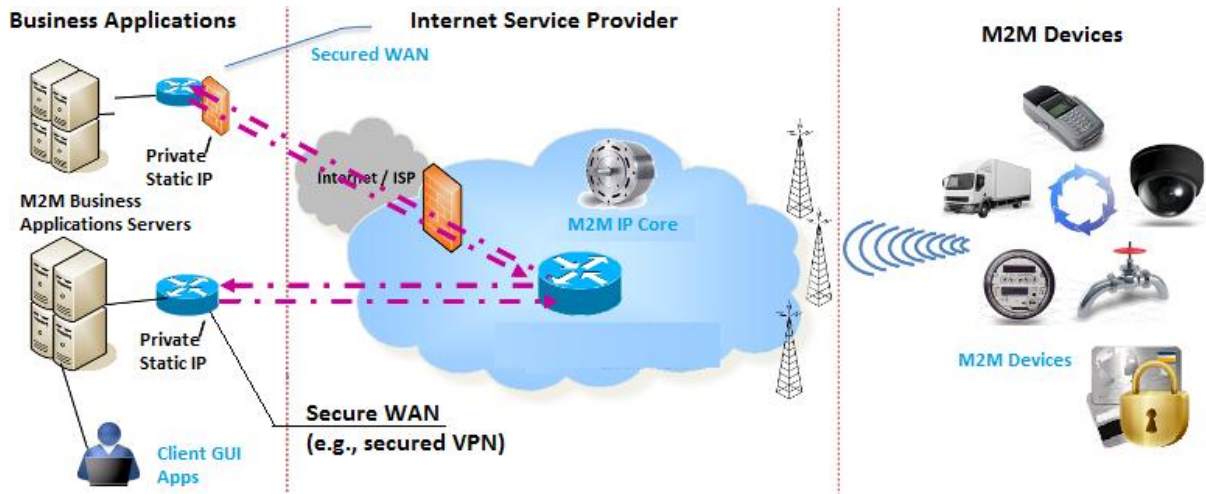

Figure 5. 5: Integration of ubiquitous systems with 5G networks

SDN paradigm can be deployed as higher layers of the protocol stack, as well as for wireless networks, such as the LR-WPANs (Low-Rate Wireless Personal Area Networks). Extending SDN to support LR-WPAN was considered impractical because these networks are highly constrained, i.e., they require numerous low-cost nodes communicating over multiple hops to cover a large geographical area, duty cycles to provide low-energy consumption to operate for long lifetimes on modest batteries. Such an approach requires cross-layer optimization, data aggregation, and low software footprint due the limited amount of memory storage and CPU processing speeds.

Wireless SDN (WSDN) remains a key challenge for the future SDN-enabled networks. The controllers should provide an appropriate module to define the rules for LR-WPAN environment. WSDN controllers should provide flexibility to support node mobility, topology discovery, self-configuration and self-organization. They also have to deal with link unreliability, robustness to the failure of generic nodes and the control node. Furthermore, although energy efficiency has been the target of diverse research works in the past, it remains an open issue that wireless Internet of things (IoT) will face. The IoT eco-system has become extremely complex and highly demanding in terms of robustness, performance, scalability, flexibility, and agility. The IoT will require new air interfaces, protocols and models optimized for short and sporadic traffic pattern. SDN should significantly reduce the cost of powering the entire network, the hardware as well as for running software. Possible solutions are for example, by shutting off IoT components when they are idle, or adapting link rates to be as minimal as possible, and even by introducing new energy-aware routing protocols [12]. In the latter case, the SDN controller collects utilization statistics for links to get visibility of flows in the network, and forward flows according to these protocols. An important consideration should be taken into account when designing energy-aware protocol is the need of the network to recover after failure while supporting automatic topology discovery at the same time.

*5.6.3 Mobile Clouds*

Mobile cloud computing is one of the technologies that are converging into a rapidly growing field of mobile and wireless network. Mobile cloud future applications in 5G will have a profound impact on almost all activities of our lives. Mobile cloud provides an excellent backend for applications on mobile devices giving access to resources such as storage, computing power etc., which are limited in the mobile device itself. The close interaction with cloud may create an environment in which mobile devices look attached locally to the cloud with low latency [13].

SDN promises an interactive solution to implement new capabilities, i.e., to enable cloud applications and services retrieve network topology, monitor the underlying network conditions (e.g., failures), and initiate and adjust network connectivity and tunneling [14]. The 5G design communication model aims to provide a global architecture with a modulator SDN layer to orchestrate the communication between the applications and services in the cloud and user's mobile terminal. Given the dynamic needs and supply of the network resource, with rich resources available in the cloud, mobile users can benefit from resource virtualization. The virtualization can abstract these dynamic mobile resources to accommodate the different requirements as elements move around within the mobile-cloud.

Despite SDN has some advantages such as resource sharing and session management, it incurs several limitations. In particular, because mobile users trigger repeatedly the embedded controller for marshalling and unmarshalling flow rules in OpenFlow messages, the overhead increases more significantly because of the limited computing capabilities and resources of mobile devices (i.e., extra memory consumption and extra latency). For example, mobile interactive applications (e.g., mobile gaming, virtual visits) require reliable connectivity to the cloud as well as low-latency and impose higher bandwidth requirements from wireless access networks to cloud service. Furthermore, mobile user's use cloud-connected devices through public and private Mobile Personal Grid (MPG), which induces multi-dimensional limitations including, dynamic mobility management across heterogeneous networks, power saving, resource availability, operating conditions), and further limits the movement of content across multiple devices and the cloud. Addressing these limitations simultaneously may increase device complexity, degrades the network performance, and causes connectivity dispersions. The key challenge for mobile clouds is how to transform physical access networks to multiple virtual and isolated networks, while maintaining and managing seamless connectivity.

## 5.7 Cooperative Cellular Networks

Another important paradigm has recently gained a lot of attention as one of the most promising technologies in next generation of wireless/cellular networking is multi-hop relay communication. Presently, cellular systems have a single direct link between the base station and the terminal. However, multi-hop networks require maintaining multi-link between multiple transmitter and receiver to form multi-path communication, so called multi-hop cooperative network. Compared to existing technology, which include mechanisms for re-transmission and multiple acknowledgments, multi-hop cooperative network can overcome these limitations by providing high-density access network. However, multi-hop cooperative network incurs several limitations and often suffer a throughput penalties since it operates in half-duplex mode and therefore introduce insufficiency of the spectrum usage.

To increase the capacity of 5G systems, SDN can provide solutions to overcome the limitations of multi-hop wireless networks [15]. Indeed, SDN can provide advanced caching techniques to store data at the edge network to reach the required high capacity of 5G systems. One way to increase per-user capacity is to make cells small and bring the base station closer to the mobile client. In cellular communications, an architecture based on SDN techniques may give operators greater freedom to balance operational parameters, such as network resilience, service performance and QoE. OpenFlow may work across different technologies (i.e., WiMAX, LTE, Wi-Fi) to provide rapid response to the subscriber mobility and avoid disruptions in the service. The decoupling between the radio network controller and the forwarding plane will enhance the performance of the base station.

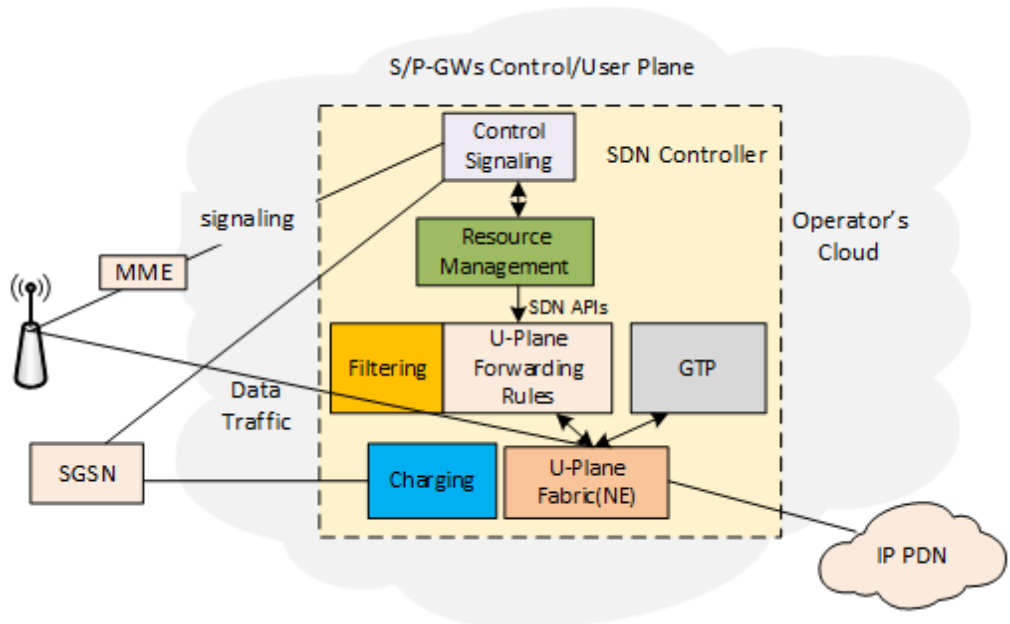

Figure 5. 6: Fully virtualized SDN-enabled cellular network

Additionally, supporting many subscribers, frequent mobility, fine-grained measurement and control, and real-time adaptation introduces flexibility, scalability and security challenges for future 5G systems architecture. SDN-enabled network devices should be able to provide scalability (i.e., increasing number of subscribers), frequent changes in user location (i.e., redirecting traffic to proxies), QoS (i.e., handling traffic with specific priority), real-time adaptation to network conditions (i.e., load balancing). Cellular SDN networks should maintain Subscriber Information Base (SIB) to translate subscribers attributes into switch rules to set up and reconfigure services flexibly.

However, the dynamic reconfiguration of a service needs a mechanism to handle notifications sent from middle boxes to the controller. Therefore, a Deep Packet Inspection (DPI) engine would be required to enable finer-grain classification based on the application (i.e., such as Web, peer-to-peer, video, and VoIP traffic). DPI also would help to support intrusion detection and prevent systems that analyze packet contents to identify malicious traffic. Likewise, cellular controller protocols would enable the control of remote virtualized resources to simplify resources and mobility management [16]. The SDN controller would enable slicing the network into multiple tenants, while enable dynamic routing, and traffic engineering, thereby easing the hand-off management, minimizing delays and packet loss may be reduced. Such a cellular SDN controller [17] (as depicted in Figure 5. 6 [16]) would implement Radio Resource Management (RRM) APIs as northbound interfaces to simplify the QoS management (i.e., admission control, resource reservation, and interference management) and the resource provisioning. The controller may be enhanced by other techniques like header compression/decompression to reduce the overhead for applications with small packet payloads (e.g., VoIP packets). Compressing these packets before transmission on low-bandwidth links substantially lowers the overhead.

Cellular networks traditionally have been hierarchical with centralized control and data structures, which require high performance, custom hardware to process, and route traffic. The distributed-control SDN model would be a key challenge for the evolution of SDN-enabled cellular network to provide high performance, cost-effective and distributed mobility management in cellular architecture. The increasing roll out of 5G technology could lead to an upsurge in SDN adoption. It seems to be possible to offload a base station the rising number of mobile clients requesting the network resources and provide load balancing strategy [18], for example, by providing multiple parallel transmissions. Furthermore, as for cloud partitioning and network slicing, it would be possible to divide the wireless traffic into several slices matching different traffic criteria, which may allow traffic isolation with respect to their patterns (i.e., VoIP, Data, Video, etc.). Such an approach may help to create virtual base stations and orchestrate the available resources among different mobile devices, thereby saving power and memory usage [19].

## 5.8 Unification of the control plane

Looking back to the development of the existing wireless communication technologies, it is easy to find that they were designed to provide new services in isolation. The future 5G cannot be defined in single type of service or isolated services. Rather, it will provide a convergent network infrastructure which integrates multiple systems integrated together. Weaving different access technologies together in a fluid fashion and creating smart gateways in transparent manner will be the goal that gives life to 5G. Leveraging SDN technologies for designing new control mechanisms and protocols for relocating functions and protocol entities will fulfill the new requirements of scale, latency, harmonization of protocol stacks between fixed and mobile (data and control planes), distributed mobility, energy efficiency and unified access/aggregation network for infrastructure simplification.

### 5.8.1 Bringing Fixed-Mobile Networking Together

The convergence of the fixed and mobile networks forms the backdrop for upgrades to the future networks. Both network infrastructures will represent the major part of investments for the network operators. The ultimate purpose is to offer better services over fixed and mobile networks with the best possible user's quality of experience, while at the same time rationalizing and sharing fixed and mobile network infrastructures.

Although some initiatives to offer some degree of converge alongside the emergence of IP-based services and IMS (IP Multimedia Subsystem), the convergence of fixed and mobile networks is a highly complex issue. Convergence is trendy word because fixed and mobile networks were developed independently from each other and based on different technologies and protocols. Convergence is also synonym to energy efficiency, because it is expected that the development of the future network will be based on ecosystem of close cooperation between fixed and mobile infrastructure. In addition to the ever-higher capacity trend, convergence of fixed and mobile networks is a highly complex issue, because it assumes certain trade-offs, so as to fully leverage off the benefits of moving different network functionalities and/or device equipment closer to each other and to different parts of the network, and should corresponds to the behavior of end-users, who wish to remain agnostic about which technical infrastructure (3GPP, Wi-Fi, DSL, fiber) they may be using.

Fixed-mobile convergence can be segmented into two concrete approaches: structural convergence and functional convergence. Structural convergence concerns sharing fixed and mobile network equipment's and infrastructures as much as possible. The functional convergence, i.e. convergence of fixed and mobile network functions, to better distribute the various functions by distinguishing those that would be centralized from those that should be more distributed. This functional convergence is enabled by NFV and SDN. The decentralization of the network functions can be provided from the mobile core network down towards the access network, such as CDN (Content Delivery Network) through the virtualization of Home Gateway to cloud functions which are mostly dictated by the traffic optimization (e.g., latency, bandwidth, etc.).

Fixed-mobile convergence is also expected to evolve different stakeholder roles: the classical network providers will continue to play there central role, but there will be other stakeholders such OTT (Over-The-Top) providers, which should be allowed to vertically integrate with content providers, or even and even the end-users themselves, will have a major involvement in the future evolving fixed-mobile ecosystem. In the context of SDN and NFV, the debate would be about exploring the network equipment that will be hosting the applications of other networks, and the functions that can migrated to the cloud.

### 5.8.2 Creating a Concerted Convergence of Packet-Optical Networks

Access technologies of future network systems would comprise various broadband transmission media such as optical fibers, millimeter-wave links, etc. The expected impact in 5G wireless communication is to contribute to the emergence of new generations of optical transport networks, to cope with the expected significant traffic growth and meet the flexibility requirements. The convergence of packet and optical networks in the future 5G system make it possible to reconfigure the optical network to support high-capacity data rate with a guaranteed end-to-end latency for on-demand applications such as NaaS (Network as a Service) [20].

Accordingly, the future generation of photonic communication would require programmable optical hardware to increase flexibility in the control plane and management plane of optical networks and enable the advent of software defined optical networking. The increased programmability of SDN creates an opportunity to address the challenges of unifying packet-optical circuit switching networks in a single converged infrastructure [21]. Unified software control of the physical layer is a key requirement for next generation 5G wireless networks. The SDN-enabled optical cross-connects would be used to demonstrate the efficiency benefits of hybrid packet-optical circuit switching architectures for dynamic management of large flows, scalability and flexibility of high-capacity service provision, in data-center applications.

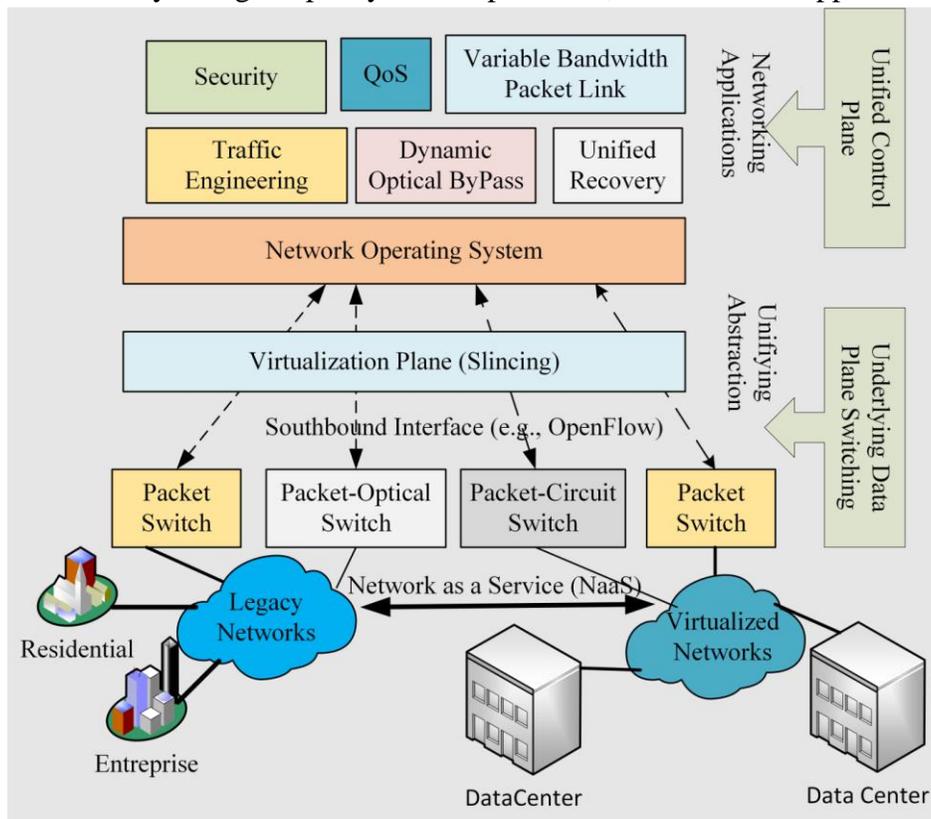

Figure 5. 7: Unifying Packet-Optical Data Plane

Similarly, future SDN-enabled 5G systems should provide a convergence framework to make it more efficient to use the network resources, for example by unifying the control and the management of these heterogeneous networks. The packet-optical networks may be unified with two of abstractions by implementing (i) a common API abstraction (flow abstraction) at the control plane, and (ii) a common-map abstraction based on a data-abstraction of a network wide common-map manipulated by a network-API. As depicted in Figure 5. 7, the common flow abstraction fits well with both networks and provides a common paradigm for control, by providing an abstraction of layers L2/L3/L4 packet headers as well as by L0/L1 circuit flows. The flow abstraction blurs the distinction between both technologies and processes them as flow of different granularity. The common-map has full visibility into both packet and circuit network devices to interconnect network applications across both packets and circuits. Full visibility allows applications to joint and optimize network functions and services across multiple layers. The network functions would be implemented as simple and extensible centralized northbound interface to hide the details of state-distribution from the applications.

## 5.9 Supporting automatic QoS provisioning

The advanced 5G network infrastructure for future Internet will include multiple heterogeneous networks that need sharing resources on all levels to meet the fast changing of traffic patterns from different services and applications. Network operators should be able to predict the various traffic patterns as functions of the services provided their networks. Service providers are evaluating implementations of storage and data traffic over a single network to meet the flexibility (e.g., the ability to accommodate short duration extra bandwidth requirements) and the efficient coexistence of multiple services [22] [23]. They have to cope with the large demands of QoS incoming from different wired and wireless devices, each with particular requirements. The QoS provisioning in the advanced SDN-enabled 5G networks to more complex and poses a real problem that need to be addressed. In particular, QoS automation should be supported at every wired and wireless technology that may share the same the network slice. Although SDN allows creating different network slices in the same network infrastructure to provide a strict QoS, performance and isolation required by across applications without interfering with traffics in other slices, however, SDN does not provide the ways for automating QoS provisioning per-application/per-service.

Indeed, one of the limitations of the OpenFlow protocol is it does not implement strict QoS in the forwarding plane. Even if some initiatives targeted implementing per-flow routing optimization to improve fined granularity for flow management [24] [25], nevertheless, resource sharing and dynamic QoS allocation was not enabled. Thus data packets will require an external tool/protocol to do so. Moreover, the current vision of SDN the QoS management is implemented at high abstraction level through the northbound interfaces. The SDN controller can map the flow requirements to the priority queues in the network device it controls and thereby reserving the network resources to individual and aggregated flows in a particular switch, but the QoS configuration cannot be done in real-time. A network administrator is required to specify the configuration of each service before the communication begins. It should install specific rules for each aggregated flows while omitting others, thereby sacrificing the fined control of services and losing the flexibility of using specific rules that match on certain packet-header fields.

In general, improving automatic QoS allocation for different and heterogeneous networks requires new methods, models and compositions to commit multiple Service Level Agreements (SLAs) end-to-end to provide a unified resulting SLA. These new mechanisms should allow services and applications evaluating SLAs in local and then be aware of all the context chaining before used in unified environments. Cloud service providers may be a good approach to follow, since virtualizing charging and security functions would improve network resiliency and availability and enforce QoS provisioning end-to-end.

## 5.10 Cognitive Network Management and Operation

The Operation and Management (OAM) of wireless mobile network infrastructure will play an important role in addressing the challenges of the future 5G system in terms of performance, constant optimization, fast-failure recovery, fast adapted in changes in the network loads, self-network organization and fast configuration. Vendor-specific OAM tools provide little or no mechanism for automatically responding to events that may occur. Besides, these tools can provide their power in small and medium networks, in contrast due to the excessive cost to deploy them in large-scale networks they will be underutilized for the future networks. Additionally, the existing OAM tools need to be individually configured and supervised by human operator which limits their flexibility. Since the network topologies are becoming more and more complex, manual configuration and deployment are getting less and less attention and are becoming impracticable. Also, the migration to high-speed networks (i.e., from 1Gbps to 10Gbps to 40Gbps) creates further scalability challenges for the future OAM tools. In particular, diagnosing the network performance and bottlenecks without visibility into the traffic characteristics introduces new complexity with regard to the consistency of the network.

Accordingly, future 5G networks should be based on common network management and operation for mobile and wireless as well as for fixed network for economic network deployment and operation. Towards the automation of network OAM tasks, network operators must grapple vendor-specific configurations to implement complex high-level interfaces to manage and monitor network policies. Advanced intelligence should be developed for realizing the future OAM tools. The intelligence of OAM requires the development of new functional and system architecture, also taking the integration of both wireless and fixed networks into account. As SDN will be the Bedrock for the future wireless/broadband networks, the OAM will be a key challenge for SDN-enabled networks.

Indeed, SDN introduces new possibilities for network management and monitoring capabilities that can improve performance, reduce the bottlenecks of the network, and enable debugging and troubleshooting of the control traffic. To avail of these possibilities, the future OAM tools should provide open and customizable interfaces to support event-driven model for SDN. SDN OAM tools should provide methodologies for the acquisition, analysis, improvement of knowledge representing the semantics and operational goals and strategies, network properties, and automated reasoning for the alignment of different network functionality at runtime. To this end, high-level, declarative management languages will be required to ensure the consistency of the network states and detect failures in real-time.

The expected impact of SDN OAM tools is to be able to scale to large-scale networks to deal with multiple controllers (i.e., in a distributed SDN model). They have to provide closed control loop functions dedicated to self-configuration, self-optimization, and self-healing. The control loop diagnostic and decision making processes need to be adapted automatically, e.g., by predicting the future actions based on the results of previous ones. This proactive capability will leverage the flexibility and programmability of the open SDN; improve their effectiveness and efficiency, thanks to cognitive processes that will enable creating more elastic network management either for the entire network or specific slices. The cognitive network management and operation approaches will develop a new management paradigm and investigate develop and verify processes, functions, algorithms and solutions that enable future 5G networks to be self-managed. The cognitive OAM will include cognitive function orchestration and coordination, and system verification for provisioning, optimization and troubleshooting.

## 5.11 Role of Satellites in the 5G networks

The 5G system had been seen an increased demands on the back-haul with an increasing numbers of HetNets and small cells. Satellites will play a major role in the extension of 5G cellular network to new area such as ships on the sea and remote land area which are not covered by cellular networks. Also, high throughput Satellite Communication (SatCom) systems will able to complement terrestrial provision in an area where it is difficult to do so with other terrestrial cell such as LTE. Indeed, integrating satellites in the future 5G networks will be seen as essential part of the terrestrial infrastructure to provide strategic solution for critical and lifesaving services. Satellites would be able used to collect and distribute data from clusters of sensors in the IoT and made them available to the terrestrial networks. Coupling SatCom systems with terrestrial cellular networks to integrate new use cases with satellites will provide a powerful new fusion enabling the innovation of services.

As SatCom systems can provide an overlay network, the integration of NVF/SDN would enable including network node functions on board satellite to save on physical sites on the ground and open up new chances to improve network resiliency, security and availability. The use of NFV and SDN in SatCom will allow networks to react on demand of the users whenever they are. They also allow the dynamic reconfiguration of the network to give users the perception of infinite capacity of their applications. Satellites would provide a wide coverage area of wireless networks to extend the dense of terrestrial cells. They can provide larger cells in heterogeneous arrangement to supply critical and emergency services take on and keep alive the network in cases of disasters. They also can be able to relieve terrestrial cells of signaling and management functions in a software defined network configuration. Satellites will be integrated to the terrestrial system to improve QoS as well as QoE to end-users.

The future SatCom systems will be able to provide intelligent traffic routing among the delivery systems, caching high capacity video to off-load the traffic from the terrestrial networks, and thereby enable saving on valuable terrestrial spectrum. In particular, one of the key drivers of 5G network architecture is the lack of spectrum that would be used for the future wireless infrastructure, so frequency sharing between mobile and satellite systems can deliver major increases in the spectrum provided both sectors. Leveraging SatCom systems with techniques like SDN, NFV, cognitive and software-defined radio can be built into future systems to allow such frequency sharing.

The extension of SDN to satellites would provide an attractive perspective for the SatCom community. By exploiting SDN/NFV satellite equipment will not be vendors-specific; instead they will be open, programmable and reconfigurable platforms. SDN and NFV are expected to offer new cost-effective services, since SatCom operators will be offering the ability to monetize on their network while offer these future/expected services. For example, the emergence of Cloud-RAN would be an enabled for virtualizing SatCom resources (i.e., ground equipment's, aerospace access infrastructure), even more the applying NFV and Cloud-RAN to SatCom paves the way towards the full virtualization of satellite head-ends, gateways/hubs and even Satellite terminals, thus entirely transforming SatCom infrastructure, enabling novel services and optimizing resource usage.

Network virtualization is considered as the key enabler for the efficient integration of the satellite and terrestrial domains. Via the unified management of the virtualized satellite and terrestrial infrastructures, fully integrated end-to-end network slices can be provided, integrating heterogeneous segments in a seamless and federated way. Additionally, the integration of satellite within the 5G future network will extend the coverage of SatCom systems to support new services such public transport service, vehicle to vehicle, surveillance with UAVs, high-definition video monitoring, localization and positioning. Moreover, non-geostationary satellites are actually investigated to achieve optimal networking and latency. Intelligent gateways can be designed to improve network resource use by providing hybridization of satellites and ADSL (Asymmetric Digital Subscriber Line) networks. Also, the virtualization can be used to provide black-box (flight data recorder) in the cloud for passenger's aircraft.

The role of satellites in the future 5G networks reveals many challenges to support flexible, programmable and secure infrastructure. As satellites will be integrated in 5G broadband networks they should enable extending the coverage of cellular backhaul, while at the same time providing enhanced user-centric QoE, cost-effective user terminals and energy efficiency. SatCom systems should continue to honor guaranteed service delivery to end-users by providing higher throughput and low latency for interactive and immersive services independently from the user location. Additionally, the integration of satellites in 5G networks will introduce new challenges regarding the spectrum sharing. Since mobile terminal will use both terrestrial connectivity as well as satellite connection, mobile receivers should support both kinds of connectivity. Thus, multi-polarized schemes are key challenge for satellite and context aware multi-user detection. Techniques like SDN, NFV and SDR (i.e., mobile terminals will have modulation and new error-control schemes that will be downloaded from the Internet on the run) are been seen as more challenging aspect of 5G networks so they should be able to provide intelligent orchestration as well as smart antenna beam forming to enable and facilitate frequency sharing between terrestrial and satellite systems.

## 5.13 Conclusion

Evolution, convergence and innovation are considered the technology routes towards 5G to meet a wide range of services and applications requirements of the information society in 2020 and beyond. To that end, a network must be designed with the future in mind, so that hardware could be abstracted and dynamically utilized through virtualization technologies, which is why a holistic SDN and NFV strategies are paramount.

The 5G network will be a combination of multi-systems, multi-technologies which need to share the frequency spectrum as well as the physical infrastructure. Nevertheless, wireless and mobile networks will pose challenging issues regarding their integration in the future 5G wireless/mobile broadband world. Leveraging SDN and NFV for supporting and improving LTE networks remains an open issue that should address the way the network functions and components will be moved a secured and virtualized cloud. SatCom system poses also challenging issues on how satellites will be integrated to the terrestrial backhaul wireless network, in such a way to provide heterogeneous segments in a seamless and federated way. Security is an open issue in SDN-enabled 5G networks as well. The programmability of SDN presents a complex set of problems facing the increasing vulnerabilities, which will change the dynamics around securing the wireless infrastructure.